\documentclass[aps,pra,twocolumn,showpacs,amsmath]{revtex4-1}
\usepackage{graphicx}
\usepackage{amssymb}

\newcommand{\bra}[1]{\langle #1 | \,}
\newcommand{\ket}[1]{\, | #1 \rangle}

\newcommand{\iim}{\mathrm{i}}
\newcommand{\be}{\begin{equation}}
\newcommand{\ee}{\end{equation}}
\newcommand{\bea}{\begin{eqnarray}}
\newcommand{\eea}{\end{eqnarray}}
\newcommand{\besa}{\begin{subeqnarray}}
\newcommand{\eesa}{\end{subeqnarray}}
\newcommand{\bean}{\begin{eqnarray*}}
\newcommand{\eean}{\end{eqnarray*}}

\begin{document}

\title{Quasi-one-dimensional scattering in a discrete model}

\author{Manuel Valiente}
\affiliation{Lundbeck Foundation Theoretical Center for Quantum System Research, Department of Physics and Astronomy, Aarhus University, DK-8000 Aarhus C, Denmark}
\author{Klaus M\o lmer}
\affiliation{Lundbeck Foundation Theoretical Center for Quantum System Research, Department of Physics and Astronomy, Aarhus University, DK-8000 Aarhus C, Denmark}
\date{\today}

\begin{abstract}
We study quasi-one-dimensional scattering of one and two particles with short-range interactions on a discrete lattice model in two dimensions. One of the directions is tightly confined by an arbitrary trapping potential. We obtain the collisional properties of these systems both at finite and zero Bloch quasi-momenta, considering as well finite sizes and transversal traps that support a continuum of states. This is made straightforward by using the exact {\it ansatz} for the quasi-one-dimensional states from the beginning. In the more interesting case of genuine two-particle scattering, we find that more than one confinement-induced resonance appear due to the non-separability of the center-of-mass and relative coordinates on the lattice. This is done by solving its corresponding Lippmann-Schwinger-like equation. We characterize the effective one-dimensional interaction and compare it with a model that includes only the effect of the dominant, broadest resonance, which amounts to a single-pole approximation for the interaction coupling constant.

\end{abstract}


\maketitle

\section{Introduction}
Low-dimensional systems have been among the dreams of theorists for decades \cite{Mattis}, since they provide a perfect playground where many analytical techniques become exact for a variety of models. For instance, we have the Bethe {\it ansatz} for the one-dimensional Bose gas with contact interactions \cite{LiebLiniger}, or bosonization for Luttinger's model \cite{LiebMattis,Luttinger}. Simple product ground-state wave functions are exact in supersymmetric one-dimensional quantum systems, with Sutherland's model \cite{Sutherland} their main representative. In addition, numerical methods such as the density-matrix renormalization group \cite{DMRG} and its recent formulation in terms of matrix product states \cite{Schollwoeck} can handle non-integrable models efficiently and accurately in one dimension. 

With the advent of ultracold atom and molecule physics \cite{BlochReview}, it is now possible to engineer effective two- and one-dimensional many-body systems by strong confinement in one or two dimensions. This possibility, combined with the great degree of control of two-body interactions thanks to magnetic Feshbach resonances \cite{Feshbach,Bauer,FeshbachReview}, has led to major experimental achievements. For example, a Tonks-Girardeau gas \cite{Girardeau} -- a one-dimensional gas of impenetrable bosons -- has been experimentally realized in an optical lattice \cite{Paredes}. In 1998, Olshanii derived how short-range interacting bosons effectively interact in the quasi-one-dimensional regime when tight confinement is applied in two of the space directions \cite{Olshanii}, and predicted the existence of a so-called confinement-induced resonance under experimentally relevant conditions. This type of resonances was used in \cite{Haller} to create and characterize a super Tonks-Girardeau gas, that is, a highly excited Bose gas state where the interactions are strongly attractive.

Confinement-induced resonances have been recently studied experimentaly in detail in \cite{Haller2}. There, Haller and co-workers found that, as the transversal trap was made anisotropic, a splitting and shift of an inelastic resonance -- only atom losses were measured -- occured, and that additional resonances appeared as anisotropy was further increased. There have been various theoretical attempts to explain the splitting of the resonance. In \cite{Drummond}, Peng {\it et al.} studied two-body s-wave collisions under anisotropic harmonic confinement, and found that in this model a shift in Olshanii's resonance does occur, but no splitting was observed. Moreover, the resonance positions are shifted from the experimental data. This issue has been beautifully resolved by Sala and collaborators in \cite{Sala}, where they showed that the splitting is due to non-separability of the center-of-mass and relative motion in two-body scattering, since in the actual experiment an optical lattice -- intrinsically anharmonic -- was used. Sala {\it et al.}'s theory and results have subsequently been corroborated in \cite{Drummond2}.

In this paper, we study one- and two-body scattering with zero-range interactions in the quasi-one-dimensional regime. The particles live in two spatial dimensions, for simplicity. To avoid unnecessary complications with the singular nature of the zero-range potential, we focus on a lattice model. An interesting by-product of our discrete model is that the center-of-mass and relative motion in the two-body problem are not separable. We are therefore able to confirm the appearance of several confinement-induced resonances and we also predict the occurrence of confinement-induced cancellation of the effective interaction. We deal here as well with some general situations that have not been considered so far: transversal confinement supporting a continuum of states and finite quasi-one-dimensional systems subjected to periodic boundary conditions. The latter case is relevant to ongoing research on ultracold atoms in ring geometries \cite{rings}.
    
\section{Potential scattering}\label{sectionpotentialscattering}

We begin by studying a single particle colliding with a zero-range potential barrier or well. This is equivalent to the two-body problem when it is separable into center-of-mass and relative coordinates, the case first studied by Olshanii \cite{Olshanii}. We first derive the relevant expressions for the scattering length and phase-shifts in the case of an infinite system, and we then obtain our results for finite systems and transversal traps supporting a continuum.

\subsection{Infinite quasi-one-dimensional space in a trap}\label{subtrap1}

We study the problem of a single particle in two spatial dimensions, where only one of the directions ($y$) is trapped, while the other ($x$) is free. At $(x,y)=(0,0)$ there is a zero-range potential of strength $U$. We use a lattice model instead of the continuum one to avoid unnecessary complications with the irregularity of the delta potential in dimensions higher than one. Moreover, quasi-1D physics has not been studied so far on a lattice and shows interesting phenomena for the non-separable two-body case, as we will see. The Hamiltonian of the system reads
\be
H=H_x+H_y + U\delta_{x,0}\delta_{y,0},\label{Hampot}
\ee
where $\delta_{a,b}$ is a kronecker delta and
\begin{align}
H_x &= -J \sum_x (\ket{x+1}\bra{x}+\ket{x}\bra{x+1}),\label{Ham1}\\
H_y &= -J \sum_y (\ket{y+1}\bra{y}+\ket{y}\bra{y+1}+V(y)\ket{y}\bra{y}),\nonumber
\end{align}
with $J>0$ the single-particle tunneling rate, $x$ and $y$ integer numbers (we set the lattice spacing $d\equiv 1$) denoting the position of the particle, and $V(y)$ is a trapping potential supporting no continuum states. The trap $V(y)$ can have any shape and strength, and we begin our analysis with a harmonic potential, $V(y)=\Omega y^2$; the properties of the spectrum and eigenfunctions of $H_y$, $\{E_n\}_{n\ge 0}$ and $\{\psi_n\}_{n\ge 0}$, have been studied in \cite{ReyHarmonic,MVDPHarmonic}.

We start by finding the one-dimensional scattering length $a$ of the system. That is, we obtain the lowest-energy scattering solution to the stationary Schr\"odinger equation $H\Psi=E\Psi$,  with eigenenergy $E=-2J + E_0$. The simplest way to proceed is by proposing the following exact {\it ansatz} 
\be
\Psi(x,y)=(|x|-a)\psi_0(y)+\sum_{n=1}^{\infty} b_n \alpha_n^{|x|} \psi_n(y), \label{Ansatz1}
\ee
where $|\alpha_n|<1$ must be satisfied for each $n$, provided $E>-2J$. At higher quasi-momenta, $|\alpha_n|$ may attain unit value for low $n$. In Eq. (\ref{Ansatz1}), $b_n$ are some expansion coefficients to be determined. Two remarks on the form of Eq. (\ref{Ansatz1}) are in order. (i) The first term is the lowest-energy scattering solution (the analog to the zero-energy solution in the continuum) in one dimension \cite{MVGeneral} multiplied by the transversal ground state; it represents the quasi-1D solution we are looking for. (ii) Every term in the infinite sum corresponds to a virtual excitation to the $n$-th trapped transversal state.. (iii) The same {\it ansatz} works for the continuum theory of \cite{Olshanii}, of course.

\begin{figure}[t]
\includegraphics[width=0.44\textwidth]{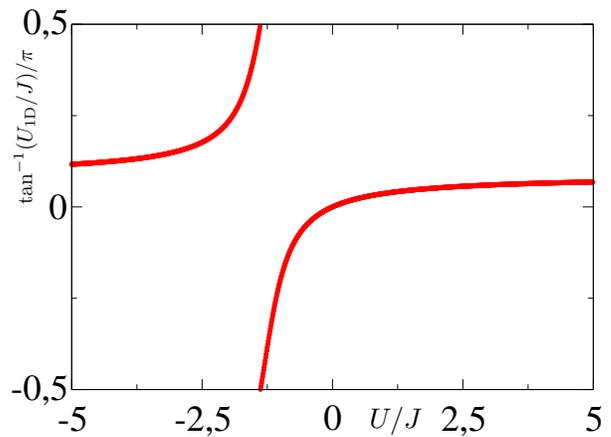}
\caption{Inverse tangent of effective interaction as a function of the bare single-particle potential strength for $\Omega/J=10^{-3}$.}
\label{fig:U1D}
\end{figure}

We introduce the {\it ansatz} (\ref{Ansatz1}) in the Schr\"odinger equation $H\Psi=E\Psi$ and, for $x\ne 0$, it is satisfied provided 
\be
-J\frac{1+\alpha_n^2}{\alpha_n}+E_n = E,\label{Alphas1}
\ee
for all $n$. The above relation expresses the fact that the energy of the virtual bound state \cite{MVGeneral} lowers the total energy to $E$ with respect to $E_n$. This point is not specific to our model, but appears in general multi-channel scattering problems involving an energy gap to coupled excitations \cite{MVMKAS}. We now use Eqs. (\ref{Ansatz1}) and (\ref{Alphas1}) in the Schr\"odinger equation at $x=0$, and project onto the transversal eigenstates $\psi_n$ \cite{Olshanii,Drummond}, obtaining
\begin{align}
&\Psi(0,0)= \frac{2J}{U\psi_0^*(0)}, \label{Psi001} \\
&b_n = U \frac{\psi_n^* (0) \Psi(0,0)}{2J\alpha_n + E - E_n}.\label{Coeff1}
\end{align}
Inserting the relations (\ref{Coeff1}) into $\Psi(0,0)$, Eq. (\ref{Ansatz1}), and equating it to Eq. (\ref{Psi001}) we find the scattering length
\be
a=-\frac{2J}{U|\psi_0(0)|^2} \left[1-U\sum_{n=1}^{\infty} \frac{|\psi_n(0)|^2}{2J\alpha_n + E -E_n}\right]\label{scatlength1}
\ee
At low energies, we can map our problem into an effective one-dimensional model. To do so, recall that the scattering length for a particle colliding with a 1D delta potential of strength $U_{\text{1D}}$ on a lattice is given by $a_{\text{1D}}=-2J/U_{\text{1D}}$ \cite{PiilMolmerNygaard,MVDP2008}. Therefore, the effective 1D potential strength $U_{\text{1D}}$ is given by
\be
U_{\text{1D}} = \frac{U|\psi_0(0)|^2}{1-U/U_{\text{CIR}}},\label{effectiveU1}
\ee
where $U_{\text{CIR}}$ is the value of $U$ at which there is a confinement-induced resonance (CIR), that is, $U_{\text{1D}}\to \infty$. It is given by
\be
1/U_{\text{CIR}}=\sum_{n=1}^{\infty}\frac{|\psi_n(0)|^2}{2J\alpha_n+E-E_n}.\label{Ucir1}
\ee
In Fig. \ref{fig:U1D} we show the effective 1D interaction strength $U_{\text{1D}}$ as a function of the bare coupling strength $U/J$ for a shallow harmonic trap with $\Omega/J=10^{-3}$. Note that the expression obtained for $U_{\text{1D}}$, Eq. (\ref{effectiveU1}), coincides with its continuum counterpart \cite{Olshanii}. They differ only in the specific, model-dependent value of $U_{\text{CIR}}$. The two limiting cases of weak and strong $U$ are also analogous: $U_{\text{1D}}=0$ for $U= 0$, while $U_{\text{1D}}\to -U_{\text{CIR}}|\psi_0(0)|^2$ when $U\to \infty$. The major difference is that the continuum CIR is found to occur at a positive (renormalized) coupling strength (in 3D), while in our case it occurs at negative values (see Fig. \ref{fig:U1D}); this has to be so, since the bare coupling constant in the continuum is brought to the opposite sign after renormalization.

We now consider the case of a finite incident momentum $k$ ($0<|k|<\pi$). The {\it ansatz} for the lowest energy solution is readily generalized by simply changing $|x|-a$ into $\cos(k|x|+\delta_k)$ in (\ref{Ansatz1}), where $\delta_k$ is the phase shift to be determined. In Eq. (\ref{Alphas1}), the energy $E$ is substituted by $E(k)=-2J\cos(k)+E_0$. After analogous manipulations to those carried out for the scattering length above, we obtain 
\be
\tan \delta_k = -\frac{U\text{csc} (k)}{2J} \frac{|\psi_0(0)|^2}{1-U/U_{\text{CIR}}(k)},\label{phaseshifts1}
\ee
where
\be
1/U_{\text{CIR}}(k) = \sum_{n=1}^{\infty}\frac{|\psi_n(0)|^2}{2J\alpha_n+E(k)-E_n}.\label{Ucirk1}
\ee
The above expression reduces to Eq. (\ref{Ucir1}) for $k=0$, so $U_{\text{CIR}}=U_{\text{CIR}}(0)$. For finite momenta, by comparison with the pure 1D solution \cite{MVDP2008,PiilMolmerNygaard}, we obtain the momentum-dependent effective interaction
\be
U_{\text{1D}}(k) =\frac{U|\psi_0(0)|^2}{1-U/U_{\text{CIR}}(k)},\label{effectiveU1k}.
\ee

There is a relevant remark to make about the above results. First, it is clear that, while at weak potential strength $U$ the effective 1D properties are dominated by the bare coupling, in the hard-core limit these are dictated by the virtual excitations in the trap, that is, the position of the confinement-induced resonance, $U_{\text{CIR}}$. This means that many-body physics in the limit of hard-core interparticle interactions will be solely determined by the virtual transitions, whenever the densities are low enough for having all $|\alpha_n|<1$ in Eq. (\ref{Alphas1}) at zero and finite relative quasi-momenta.


\subsection{Infinite quasi-one-dimensional space in a trap supporting a continuum}

So far, we have dealt with the case of a transversal trapping potential that supports only bound states. Such a situation is obviously ideal, and we here address this issue within the potential scattering approach.

We consider a single-particle described by Hamiltonian (\ref{Hampot}), with $H_x$ and $H_y$ defined by Eq. (\ref{Ham1}), and $V(y)$ a transversal potential well such that $H_y$ supports at least one and at most a finite number of bound states. For convenience, we assume that $V$ is symmetric ($V(y)=V(-y)$) and monotonic ($V(|y|+1)\ge V(|y|)$), and we add an energy off-set so as to render it positive, $V(y)=\tilde{V}(y)+V_0$, with $V_0 = -\tilde{V}(0)>0$. We will consider $V$ to have an arbitrary but finite range $R\ge 0$, that is, $V(|y|>R)=V_0$.

The symmetric bound states of the transversal Hamiltonian $H_y$ are readily obtained by calculating the roots of a certain polynomial of degree $2R + 1$ if $R\ge 1$ \cite{MVGeneral} or of degree 2 for the zero-range case \cite{MVGeneral,MVDP2008}. The antisymmetric states do not contribute to the scattering length/amplitude. The scattering states, with their corresponding phase-shifts are also easily calculated \cite{MVGeneral}. 

The effective quasi-1D scattering lengths and phase shifts have the same expressions as for the trapping case, Eqs. (\ref{scatlength1}) and (\ref{phaseshifts1}), while special care has to be taken in the sums of Eqs. (\ref{Ucir1}) and (\ref{Ucirk1}). First of all, notice that if $H_y$ had a purely continuous spectrum, the quasi-1D picture would break down and the sum in Eq. (\ref{Ucirk1}) would not be well-defined; in such case genuine two-dimensional scattering solutions must be considered using, e.g. the Lippmann-Schwinger equation in quasi-momentum space. Fortunately, this situation happens only if $V\equiv 0$, and we need not worry about repulsively-bound states \cite{MVDP2008,PiilMolmerNygaard,Winkler} above the transversal continuum if we choose $V(R) < V_0$. 

The symmetric scattering states of $H_y$ have the form
\begin{equation}
\psi_q(y)=\left\{
  \begin{array}{rl}
    &\frac{1}{\sqrt{N}} \cos(q|y|+\theta_q)  \text{ if } |y| \ge R, \\
     &\frac{1}{\sqrt{N}} \phi_q(y)  \text{ if } |y|<R.\label{psiq1}
    \end{array} \right.
\end{equation}
In the above equation, $q$ is a quasi-momentum, $\theta_q$ are the phase shifts, $\phi_q$ is the inner part of the scattering wave function to be determined, and $N$ is a quantization length unit which ensures normalization for $N\gg R$.

In order to be able to perform sums over continuum states, we need their density of states. Applying the boundary condition $\psi_q(L)=0$, with $2L+1=N$, to Eq. (\ref{psiq1}), we obtain the usual discrete set of allowed momenta $q=q_n=(2n+1)\pi/2L - \theta_q/L$. Following a procedure parallel to that of ref. \cite{Huang} (see chapter 10), we obtain the density of states $g(q)$,
\be
g(q) = \frac{L}{\pi} + \frac{1}{\pi} \frac{\partial \theta_q}{\partial q}.
\ee
We express the discrete sum of any function $F$ as
\be
\sum_q F(q) = \sum_q F(q) g(q) \Delta q \to \int_{-\pi}^{\pi} dq F(q) g(q),
\ee
where we have used that $g(q) \Delta q = 1$ and we have assumed that the limit $L\to \infty$ can be taken to replace the sum by an integral over the Brillouin zone. We then obtain the relevant integral over continuum states
\begin{align}
\mathcal{S}(k)\equiv&\sum_q \frac{|\psi_q(0)|^2}{E(k)-E(q)+(E_0-V_0)+2J\alpha_q} \rightarrow \nonumber \\
&\frac{1}{2\pi} \int_{-\pi}^{\pi} dq \frac{|\phi_q(0)|^2}{E(k)-E(q)+(E_0 -V_0) +2J\alpha_q} \nonumber\\ + &\lim_{L\to \infty} \frac{1}{2\pi L}\int_{-\pi}^{\pi} dq \frac{|\phi_q(0)|^2\partial \theta_q /\partial q}{E(k)-E(q)+(E_0 -V_0) +2J\alpha_q}  \label{sumcont1}
\end{align}
In the above equation, $E(k)=-2J\cos(k)+E_0$ is the energy of the target quasi-1D state ($k$ is the one-dimensional quasi-momentum), and $\alpha_q$ satisfy Eq. (\ref{Alphas1}) with their corresponding energies changed as $E_n \to E(q)+V_0-E_0$. The last term in Eq. (\ref{sumcont1}) is kept since it may contribute in the case of a sharp resonance ($\partial_q \theta_q =\infty$). If a sharp resonance at $q=q_0$ contributes to (\ref{sumcont1}), then $\partial_q \theta_q \propto L \delta(q-q_0)$ for large $L$ \cite{Huang}.

We now have all we need for the solution to our problem, so that the position $U_{\text{CIR}}$ of the CIR is given by
\be
1/U_{\text{CIR}}(k) = \sum_{n=1}^{\mathcal{M}-1}\frac{|\psi_n(0)|^2}{2J\alpha_n+E(k)-E_n} + \mathcal{S}(k), \label{Ucirk2}
\ee
with $\mathcal{S}$ given by Eq. (\ref{sumcont1}), and $\mathcal{M}$ the number of bound states supported by the trap.

\begin{figure}[t]
\includegraphics[width=0.44\textwidth]{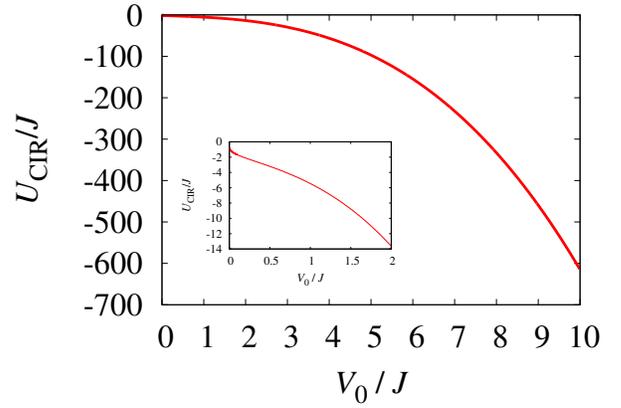}
\caption{Position of the CIR as a function of the trap depth for a model trap supporting a single bound state and a continuum (see text). In the inset, an augmented portion of the figure shows the resonant positions before the transversal trap reaches the extreme strong-coupling limit.}
\label{fig:UCIR-continuum}
\end{figure}

To illustrate how continuum states of $H_y$ are included, we consider the simplest possible transversal potential $V(y)=V_0-V_0\delta_{y,0}$, with $V_0>0$, for which the above expressions can be obtained analytically. In this case, the number of bound states is $\mathcal{M}=1$. We calculate the effective quasi-1D scattering length, so we set $E=E(0)=-2J+E_0$, and we have \cite{MVDP2008,PiilMolmerNygaard} $E_0 = -\sqrt{V_0^2+4J^2}+V_0$. The phase shifts of transversal scattering states are given by $\tan \theta_q = V_0/(2J\sin q)$, and therefore there are no sharp resonances. The values of $\alpha_q$ are given by
\be
\alpha_q = f(q,V_0/J)-\sqrt{f(q,V_0/J)^2 -1},
\ee
with $f(q,r)\equiv 1-\cos q +\sqrt{(r/2)^2+1}$, and hence satisfy $0<\alpha_q<1$ for all $q$ for $V_0 > 0$. From Eq. (\ref{sumcont1}) we then have
\be
\mathcal{S}(0) = \frac{1}{2\pi} \int_{-\pi}^{\pi} dq \frac{1}{1+\frac{V_0^2}{(2J\sin q)^2}} \frac{1}{2J(\cos q  - 1) + 2J\alpha_q},
\ee
which is finite, as we expected. Note that, in this very simple model, it holds that $1/U_{\text{CIR}} = \mathcal{S}(0)$. In Fig. \ref{fig:UCIR-continuum} we show the resonance position $U_{\text{CIR}}$ as a function of the model trap strength $V_0$.

\subsection{Finite quasi-one-dimensional space in a trap}

We here construct quasi-1D scattering states with periodic boundary conditions (PBC) in the $x$ direction. This represents the case of a particle in a ring trap of circumference $L$ on the surface of a very long cylinder wrapped around the $y$-axis, interacting with a zero-range potential at $(x,y)=(0,0)$.  The following {\it ansatz} represents the exact quasi-1D scattering states (with energy $E=-2J\cos(k)+E_0$) subjected to PBC
\begin{align}
\Psi(x,y)&=\cos(k(|x|-L/2)) \psi_0(y) \label{toyAnsatz}\\
 &+ \sum_{n=1}^{\infty} b_n \left[\alpha_n^{|x|}+\alpha_n^{L}\alpha_n^{-|x|}\right]\psi_n(y),\nonumber
\end{align}
where $x$ is defined mod $L$, $b_n$ are expansion coefficients, and $\alpha_n$ satisfy the relations $-J(\alpha_n+1/\alpha_n)+E_n = -2J\cos(k)+E_0$, with $|\alpha_n|<1$ for all $n$. Note that in this case no scattering length can be defined, since for any non-zero interaction strength $U$ and $L<\infty$, $k=0$ is not an allowed quasi-momentum. We remark that the phase shift has been fixed so that the state satisfies PBC, and is given by the same expression as that in the Lieb-Liniger model \cite{LiebLiniger} particularized to a lattice; note that we need one further condition to determine the value of $k$, given below.

\begin{figure}[t]
\includegraphics[width=0.44\textwidth]{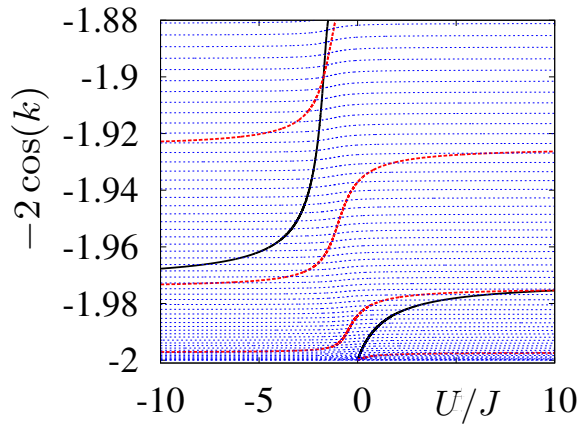}
\caption{Allowed scattering energies $-2J\cos(k)$ as functions of the single-particle interaction strength $U/J$ for a finite 1D space with periodic boundary conditions, of length $L=10$ (black solid lines), $L=50$ (red thick dashed lines) and $L=1000$ (blue thin dashed lines), for $\Omega/J=10^{-3}$.}
\label{fig:finite-energies}
\end{figure}

After introducing $\Psi$ into the Schr\"odinger equation $H\Psi=E\Psi$, with $E=-2J\cos(k)+E_0$, and following a procedure parallel to the previous subsections we obtain a nonlinear equation for the ground-state momentum $k$
\begin{align}
& 2J\sin(k)\tan(kL/2)= \\ 
& \frac{U|\psi_0(0)|^2}{1+U\sum_{n=1}^{\infty} \frac{|\psi_n(0)|^2}{(E_n-E)(1+\alpha_n^2)-2J(\alpha_n+\alpha_n^{L-1})} }. \label{finitephaseshift}\nonumber\end{align}
Allowed quasi-1D energies (with the zero mode of the trap $E_0$ as a reference) for different system sizes are plotted in Fig. \ref{fig:finite-energies}, where finite-size effects are apparent. A CIR occurs every time the energy crosses the ``fermionized'' energy, that is, the energy of a non-interacting, antisymmetric quasi-1D state ($k=(2n+1)\pi/L$), which can only happen for $U<0$.

A continuum of transversal states is again dealt with by Eqs. (\ref{sumcont1}) and (\ref{Ucirk2}), where we now have to replace the terms $2J\alpha_n$ and $2J\alpha_q$ by $2J(\alpha_n+\alpha_n^{L-1})$ and $2J(\alpha_q+\alpha_q^{L-1})$, respectively. 

\section{Two-body scattering}

In the system we are considering in this article, two-particle collisions cannot be studied via separation of the CM and relative coordinates. It is therefore of interest to obtain the quasi-1D properties of two-body scattering, for which non-separability effects may be relevant. These have proved to be fundamental \cite{Sala} to explain the experimental results of Haller {\it et al.} \cite{Haller2}.

We consider the quasi-1D space to be an infinitely long line, with the dynamics governed by the following Hamiltonian
\be
H = H_{x_1}+H_{x_2}+H_{y_1}+H_{y_2} + U\delta_{x_1,x_2}\delta_{y_1,y_2},
\ee
where $(x_i,y_i)$, $i=1,2$, represents the position of particle $i$ in two dimensions, $H_{x_i}$, $H_{y_i}$ are as defined in Eq. (\ref{Ham1}), and $U$ is the two-body interaction strength. 

We solve the stationary Schr\"odinger equation $H\Psi = E \Psi$, with $\Psi$ corresponding to a quasi-1D scattering state. The exact symmetric {\it ansatz} for two bosons (or two spin-singlet fermions) now has the form
\begin{align}
\Psi(x_1,y_1;x_2,y_2)&=e^{\iim K X} \cos(k|x_1-x_2| + \delta_k) \psi_0(y_1) \psi_0(y_2) \nonumber \\ 
&+ e^{\iim K X} \mathcal{F}(x_1,y_1;x_2,y_2),\label{Ansatztwobody1}
\end{align}
where $X = (x_1+x_2)/2$, $K$ is the (conserved) total quasi-momentum in the $x$-direction, and where $\mathcal{F}$ represents the closed multi-channel scattering part of the wave function,
\be
\mathcal{F}(x_1,y_1;x_2,y_2)=\sum_{n_1 \le n_2} b_{n_1,n_2} \alpha_{n_1,n_2}^{|x_1-x_2|}\hat{S}[\psi_{n_1}\psi_{n_2}](y_1,y_2).
\ee
In the above, $\hat{S}$ is a symmetrization operator \cite{footnotehatS}, and the sum excludes $(n_1,n_2)=(0,0)$. After introducing (\ref{Ansatztwobody1}) into the Schr\"odinger equation, we obtain the condition for $\alpha_{n_1,n_2}$ so that the {\it ansatz} is asymptotically correct,
\be
-J_K \frac{1+\alpha_{n_1,n_2}^2}{\alpha_{n_1,n_2}} = E - E_{n_1} - E_{n_2},
\ee
with $J_K = -2J\cos(K/2)$ the collective tunneling rate \cite{MVDP2008,PiilMolmerNygaard}. Following a methodology analogous to that for potential scattering, we obtain the expansion coefficients
\be
b_{n_1,n_2} = \frac{U \mathcal{I}_{n_1,n_2}}{E+2J_K \alpha_{n_1,n_2} - E_{n_1}-E_{n_2}},
\ee
where the (unknown) functions $\mathcal{I}$ are defined as
\be
\mathcal{I}_{n_1,n_2}\equiv \sum_{y=-\infty}^{\infty} \hat{S}[\psi^*_{n_1}\psi^*_{n_2}](y,y)\Psi(0,0;y,y).
\ee
The phase shift is given by the equation
\be
2J_K \sin k \sin \delta_k = -U \mathcal{I}_{0,0}.
\ee
Defining now the overlaps between different non-interacting states in the trapped direction as
\be
\mathcal{R} (n_1,n_2;m_1,m_2) = \sum_{y=-\infty}^{\infty} \hat{S}[\psi^*_{n_1}\psi^*_{n_2}](y,y)  \hat{S}[\psi_{m_1}\psi_{m_2}](y,y),
\ee
the final non-linear system of equations in the unknowns $\mathcal{I}_{n_1,n_2}$ is found to be
\begin{align}
\mathcal{I}_{n_1,n_2} &= \mathcal{R}(n_1,n_2;0,0) \sqrt{1 - \left(\frac{U \mathcal{I}_{0,0} }{2J_K \sin k}\right)^2 }\nonumber \\
&+U\sum_{m_1\le m_2} \frac{ \mathcal{R} (n_1,n_2;m_1,m_2)}{E+2J_K\alpha_{m_1,m_2} - E_{m_1} - E_{m_2}} \mathcal{I}_{m_1,m_2},\label{Eqntwobodyfinitek}
\end{align}
where the sum is restricted, as before, to $(m_1,m_2)\ne (0,0)$.

At very low momenta, the effective 1D scattering length dominates the physics and, to calculate it, we only have to replace $\cos(k|x|+\delta)$ by $|x|-a$ in Eq. (\ref{Ansatztwobody1}). The system of equations we obtain now reads
\begin{align}
\mathcal{I}_{n_1,n_2}&=\mathcal{R}(n_1,n_2;0,0)\label{Eqntwobodyscatlength}\\
&+ U\sum_{m_1\le m_2} \frac{\mathcal{R}(n_1,n_2;m_1,m_2)}{E+2J_K\alpha_{m_1,m_2}-E_{m_1}-E_{m_2}}\mathcal{I}_{m_1,m_2}.\nonumber
\end{align}
All quantities in the above equation have the same meaning as for the non-zero quasi-momentum case, with the sum excluding $(m_1,m_2)=(0,0)$, and the relevant component being now $\mathcal{I}_{0,0}=U_{\text{1D}}/U=2J_K/aU$. Note that Eq. (\ref{Eqntwobodyscatlength}) is a system of linear equations, in contrast with Eq. (\ref{Eqntwobodyfinitek}). Eq. (\ref{Eqntwobodyscatlength}) resembles the Lippmann-Schwinger (LS) equation-- although with no singularities in the ``integrand'' -- if the following identifications are made: (i) $(n_1,n_2)\equiv \mathbf{q}_n$ serves as a {\it discrete} momentum state, (ii) $U\mathcal{R}(\mathbf{q_n};\mathbf{q_m})\equiv \mathcal{V}(\mathbf{q_n};\mathbf{q_m})$ is an effective potential, (iii) $1/(E+2J_K\alpha_{\mathbf{q}_m}-E_{m_1}-E_{m_2})$ plays the role of a non-interacting Green's function $G^0$, and (iv) $U\mathcal{I}_{\mathbf{q}_n}\equiv T(\mathbf{q}_n,0)$ represents the T-matrix. We can formally write Eq. (\ref{Eqntwobodyscatlength}) as $T=\mathcal{V} + \mathcal{V}G^0T$; in particular, this analogy allows us to write its Born series as 
\be
T = \mathcal{V} + \mathcal{V}G^0\mathcal{V} + \mathcal{V}G^0\mathcal{V}G^0\mathcal{V}+\ldots,\label{BornSeries}
\ee
which is ensured to converge whenever $U$ is smaller in magnitude than the first (smallest in magnitude) CIR coupling $U_{\text{CIR}}^{1}$. 

\begin{figure}[t]
\includegraphics[width=0.44\textwidth]{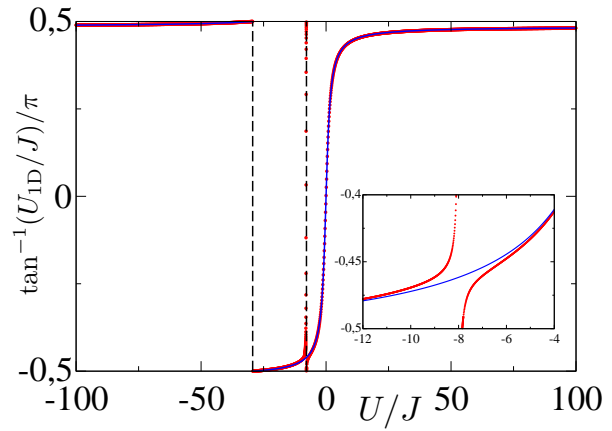}
\caption{Inverse tangent of effective interaction as function of the bare two-body interacting strength for the two-state trap model with $V=J$. Red dots represent exact results, blue solid line corresponds to single-pole approximation (see text), and dashed vertical lines mark the CIR positions. Inset: zoomed-in region near $U^{(1)}_{\text{CIR}}$, with axes labels the same as for main figure.}
\label{fig:ejriej}
\end{figure}

We will see that, due to the non-separability of the CM and relative coordinates in the trap channels, there are more than one CIR and, interestingly, there are true quasi-1D resonances ($a\to \infty$). In order to understand these, it is instructive to first study an easier problem in which the trap has only two states. This can be modelled by choosing $V(y)=2V\delta_{y,1}$ -- a discrete, transversal step function -- with open boundary conditions in the $y$-direction, $\psi_n(y=-1)=\psi_n(y=2)=0$. With this choice, both the effective Green's function and potential in the LS equation $T=\mathcal{V}+\mathcal{V}G^0T$ can be calculated analytically, and its numerical solution is trivial. In Fig. \ref{fig:ejriej}, we show $U_{1D}$ as a function of $U$ and the position of the confinement-induced resonances for a fixed value of $V/J=1$. There, we see that two CIRs and one true quasi-1D resonance are present. The rightmost resonance at $U=U^{(1)}_{\text{CIR}}$ is very sharp, while the other resonance at $U=U^{(2)}_{\text{CIR}}$ is wide and its shape is similar to that of potential scattering or, for what matters, of the case of separable CM and relative coordinates. Based on the knowledge gained in Sect. \ref{sectionpotentialscattering}, we may expect that, approximately, the position of the leftmost resonance is given by $U^{(2)}_{\text{CIR}}\approx -\lim_{U\to \infty} U_{1D} / \mathcal{R}(0,0,0,0)$, since the jump in $U_{\text{1D}}$ from the right to the left (or viceversa) of $U^{(1)}_{\text{CIR}}$ is rather small. To test this hypothesis, we start from the analogous expression to Eq. (\ref{effectiveU1}) for the two-body case, which we may call single-pole approximation (SPA),
\be
U_{1D}\approx  \frac{U\mathcal{R}(0,0,0,0)}{1-U/U^{(2)}_{\text{CIR}}}.\label{SPA1}
\ee
We expand the above expression for $U\gg U^{(2)}_{\text{CIR}}$, and perform a least squares fit $U_{\text{1D}}\approx c_1 + c_2/U$ to the numerically calculated effective interaction. Equating the SPA (\ref{SPA1}) to zero-th and first orders to the fit, we obtain
\be
U_{\text{CIR}}^{(2)}\approx -\frac{c_1}{R(0,0,0,0)} = -\sqrt{\frac{-c_2}{R(0,0,0,0)}}.\label{c1c2}
\ee
Since the SPA of Eq. (\ref{SPA1}) is not exact, the last equality in the above equation is not satisfied, but the difference between both sides of the equality provides a good test for the validity of the SPA, and a qualitative error estimation. We have done a least squares fit from $U/J=-1000$ to $U/J=-900$, using 50 equidistant points; the trapping strength is set to $V/J=1$. The results we obtain are $c_1 = 21.57J$, $c_2 = -661.22J^2$, and we have $R(0,0,0,0)=3/4$. With $c_1$ ($c_2$), from Eq. (\ref{c1c2}) we obtain a value of $U_{\text{CIR}}^{(2)}/J\approx -28.76$ ($-29.69$). Qualitatively, we locate the resonance position at $\approx -29.2J$, with an error of $O(J)$. It compares favorably with the exact location $U_{\text{CIR}}^{(2)}=-29.35J$ calculated with the determinant method ($\text{det}(\mathcal{V}G^0-1)=0$). In Fig. \ref{fig:ejriej}, we plot the SPA approximation together with the exact results, showing that they are in good agreement everywhere except in the immediate neighborhood of the sharp resonance at $U=U^{(1)}_{\text{CIR}}$, where evidently the SPA fails. 

\begin{figure}[t]
\includegraphics[width=0.44\textwidth]{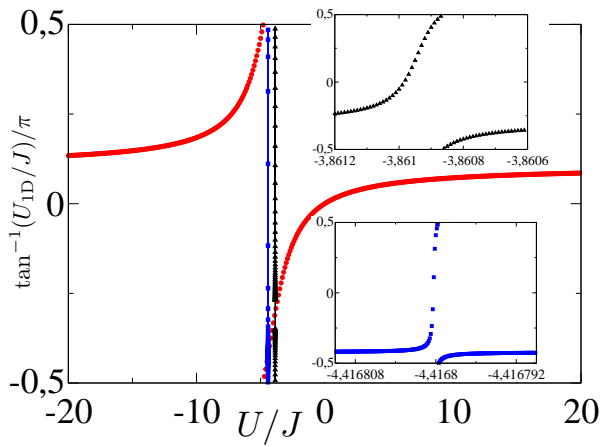}
\caption{Inverse tangent of effective interaction as function of the bare two-body interacting strength for a transversal harmonic potential with $\Omega/J=10^{-3}$ and $K=0$. Red dots, blue squares and black triangles are calculated with different steps in $U/J$ to capture the width of the three resonances. Lines are guides for the eye. Insets: zoomed-in regions near the narrow CIRs, with axes labels the same as for main figure.}
\label{fig:Uuu}
\end{figure}

We now turn to the more interesting case of a quasi-1D two-body collision under transversal harmonic confinement, $V(y) = \Omega y^2$. We solve Eq. (\ref{Eqntwobodyscatlength}) numerically and, doing so, we need an upper cut-off $N_c$ in the number of single-particle trapped states. To obtain well-converged results, we calculate $\mathcal{I}_{0,0}=U_{1D}/U$ using its Born series, Eq. (\ref{BornSeries}), to order 100 for different values for $N_c$, in the region where the Born series is well-defined. We choose $N_c$ in such a way that the Born series offers converged results (not to be confused with the convergence of the Born series itself, which is granted), and compare it with the exact numerical solution of Eq. (\ref{Eqntwobodyscatlength}). The advantage of using the Born series for this purpose is that it is capable of handling much larger system sizes than the exact solution more efficiently when it converges. Moreover, for very low values of $\Omega/J$, it is the only method -- together with resummation techniques if necessary -- that may overcome finite-size effects. In Fig. \ref{fig:Uuu}, we show the effective interaction $U_{\text{1D}}$ as a function of the bare interaction strength for a shallow harmonic trap ($\Omega/J = 10^{-3}$), which can be considered as almost separable into CM and relative coordinates \cite{MVDPHarmonic}. The cut-off $N_c=41$ is chosen since results are already well-converged. In this situation, we observe that a broad resonant profile is present, almost unaltered by two very narrow CIRs that exist due to non-separability of the coordinates. Around these resonances, the effective interaction does cross zero and therefore, in contrast to the separable case, true 1D resonances are found. Since, of the three resonances observed, two are very narrow and do not seem to contribute much to the ``overall'' shape of the red curve in Fig. \ref{fig:Uuu}, the SPA approximation, Eq. (\ref{SPA1}) with $U_{\text{CIR}}^{(2)}$ replaced by $U_{\text{CIR}}^{(3)}$, should give a good estimate of the broad resonance position. Proceeding in the same way as for the two-state trap above, we find $U_{\text{CIR}}^{(3)}/J\approx -4.8$ with an error of $O(10^{-1})$, which indeed compares very favourably with the exact value of $U_{\text{CIR}}^{(3)}/J = -4.792$. 

\begin{figure}[t]
\includegraphics[width=0.44\textwidth]{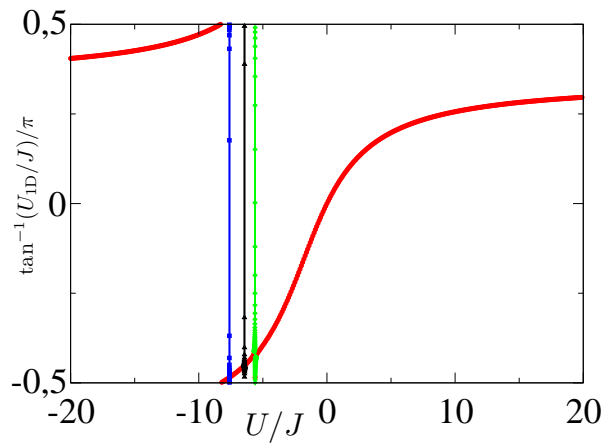}
\caption{Inverse tangent of effective interaction as function of the bare two-body interacting strength for a transversal harmonic potential with $\Omega/J=10^{-1}$ and $K=0$. Red dots, blue squares, black triangles and green diamonds are calculated with different steps in $U/J$ to capture the width of the four resonances. Lines are guides for the eye.}
\label{fig:Uuu2}
\end{figure}

If we choose a higher trapping frequency, therefore making the separability assumption less rigorous, we observe more narrow resonances, although some of them are much wider than in the almost separable case. In Fig. \ref{fig:Uuu2}, we plot the effective interaction as $U/J$ is varied for a trap strength of $\Omega/J = 10^{-1}$. The cut-off $N_c=21$ already gives converged results. In this case the SPA approximation gives, for the broad resonance position a value of $U_{\text{CIR}}^{(4)}/J\approx -8.3$ with an error of $O(10^{-1})$, while the exact position is $U_{\text{CIR}}^{(4)}/J= -8.286$.

We note that a continuum of transversal states and finite quasi-one-dimensional space for the two-body problem are dealt with in the same way as in the potential scattering approach of Sec. \ref{sectionpotentialscattering}. 

\section{Conclusions}
In this paper, we have studied quasi-one-dimensional scattering in two dimensions with one of the directions tightly confined. We have used, instead of a continuum theory, a lattice model, and seen that single-particle potential scattering under transversal harmonic trapping is analogous to its continuum counterpart. We have then considered generalized situations in which the system is finite or the transversal confinement supports a continuous spectrum. For the two-body problem, we have derived the equation governing the scattering properties of the system, and this resembles a usual Lippmann-Schwinger equation which can be solved via standard methods from scattering theory. The center-of-mass and relative coordinates cannot be separated, and therefore our model accounts for the existence of more than one confinement-induced resonance. Moreover, zero effective quasi-one-dimensional interaction strengths can be achieved, as opposed to the free-space, separable case. Our results are relevant for current on-going experimental research on dilute ultracold gases in tight-binding, anisotropic optical lattices, where these systems are extended over many lattice sites in the quasi-free direction. 

There are still many open problems in reduced-dimensional systems. Generalization and application of our results in a mean-field theory, e.g. in a transversal trap with two bound states, seems to be a natural extension of the present work. We also note that an exact treatment of the bosonic and fermionic three-body problems may show modifications on the position and properties of two-body resonances, and may as well need genuine three-body effective interactions for their correct dimensional reduction

\begin{acknowledgments} 
MV is grateful to S. Sala, A. Saenz, and especially to P.-I. Schneider for many useful discussions and feedback. Work by MV was supported by a Villum Kann Rasmussen block scholarship.
\end{acknowledgments}


\begin{thebibliography}{99}
\bibitem{Mattis}
D.C. Mattis, {\it The Many-Body Problem: An Encyclopedia of Exactly Solved Models in One Dimension} (World Scientific, 1993).

\bibitem{LiebLiniger}
E.H. Lieb and W. Liniger,
Phys. Rev. {\bf 130}, 1605 (1963).

\bibitem{Luttinger} 
J.M. Luttinger,
J. Math. Phys. {\bf 4}, 1154 (1963).

\bibitem{LiebMattis}
D.C. Mattis and E.H. Lieb,
J. Math. Phys. {\bf 6}, 304 (1965).

\bibitem{Sutherland}
B. Sutherland,
J. Math. Phys. {\bf 12}, 246 (1971).

\bibitem{DMRG}
S.R. White,
Phys. Rev. Lett. {\bf 69}, 2863 (1992). 

\bibitem{Schollwoeck}
U. Schollw\"ock,
Ann. Phys. {\bf 326}, 96 (2011).

\bibitem{BlochReview}
I. Bloch, J. Dalibard and W. Zwerger,
Rev. Mod. Phys. {\bf 80}, 885 (2008). 

\bibitem{Feshbach}
H. Feshbach,
Ann. Phys. (N.Y.) {\bf 5}, 357 (1958).

\bibitem{Bauer}
D.M. Bauer {\it et al.},
Nature Phys. {\bf 5}, 339 (2009).

\bibitem{FeshbachReview}
C. Chin, R. Grimm, P. Julienne and E. Tiesinga,
Rev. Mod. Phys. {\bf 82}, 1225 (2010). 

\bibitem{Girardeau}
M. Girardeau,
J. Math. Phys. {\bf 1}, 516 (1960).

\bibitem{Paredes}
B. Paredes {\it et al.},
Nature {\bf 429}, 277 (2004).
 
\bibitem{Olshanii}
M. Olshanii,
Phys. Rev. Lett. {\bf 81}, 938–941 (1998).

\bibitem{Haller}
E. Haller {\it et al.},
Science {\bf 325}, 1224 (2009).

\bibitem{Haller2}
E. Haller {\it et al.},
Phys. Rev. Lett. {\bf 104}, 153203 (2010).

\bibitem{Drummond}
S.-G. Peng {\it et al.},
Phys. Rev. A {\bf 82}, 063633 (2010).

\bibitem{Sala}
S. Sala, P.-I. Schneider and A. Saenz,
e-print arXiv:1104.1561 (2011).

\bibitem{Drummond2}
S.-G. Peng {\it et al.},
e-print arXiv:1107.2725 (2011).

\bibitem{rings}
A. Ramanathan {\it et al.},
Phys. Rev. Lett. {\bf 106}, 130401 (2011). 

\bibitem{ReyHarmonic}
A.M. Rey, G. Pupillo, C.W. Clark and C.J. Williams,
Phys. Rev. A {\bf 72}, 033616 (2005). 

\bibitem{MVDPHarmonic}
M. Valiente and D. Petrosyan,
Europhys. Lett. {\bf 83}, 30007 (2008). 

\bibitem{MVGeneral}
M. Valiente,
Phys. Rev. A {\bf 81}, 042102 (2010). 

\bibitem{MVMKAS}
M. Valiente, M. K\"uster and A. Saenz,
Europhys. Lett. {\bf 92}, 10001 (2010).

\bibitem{MVDP2008}
M. Valiente and D. Petrosyan,
J. Phys. B {\bf 41}, 161002 (2008). 

\bibitem{PiilMolmerNygaard}
N. Nygaard, R. Piil and K. M\o lmer,
Phys. Rev. A {\bf 78}, 023617 (2008).

\bibitem{Winkler}
K. Winkler {\it et al.}, 
Nature {\bf 441}, 853 (2006).

\bibitem{Huang}
K. Huang,
{\it Statistical Mechanics}, 2nd Ed. (Wiley \& Sons, 1987).

\bibitem{footnotehatS}
$\hat{S}[f g] (z,t) = 2^{-1/2} [f(z) g(t)+f(t) g(z)]$ if $f\ne g$, and $\hat{S}[f f] (z,t) = f(z) f(t)$.

\end{thebibliography}
\end{document}